# Mapping Cloud Computing onto Useful e-Governance

Ajay Prasad
Department of CSE
Sir Padampat
Singhania University, Udaipur,
INDIA
ajayprasadv@gmail.com

Sandeep Chaurasia
Department of CSE
Sir Padampat
Singhania University, Udaipur,
INDIA
Sandeep.chaurasia@spsu.ac.in

Arjun Singh
Department of CSE
Sir Padampat
Singhania University, Udaipur,
INDIA
vitarjun@gmail.com

Deepak Gour
Department of CSE
Sir Padampat
Singhania University, Udaipur,
INDIA
deepak.gour@spsu.ac.in

*Abstract*— Most of the services viewed in context to grid and cloud computing are mostly confined to services that are available for intellectual purposes. The grid or cloud computing are large scale distributed systems. The essence of large scale distribution can only be realized if the services are rendered to common man. The only organization which has exposure to almost every single resident is the respective governments in every country. As the size of population increases so the need for a larger purview arises. The problem of having a large purview can be solved by means of large scale grid for online services. The government services can be rendered through fully customized Service-oriented Clouds. In this paper we are presenting tight similarities between generic government functioning and the service oriented grid/cloud approach. Also, we will discuss the major issues in establishing services oriented grids for governmental organization.

I. INTRODUCTION

The analogy of electricity grid in compute grids is appropriately understood by the fact that both grids are mainly meant for supplying vastly distributed resources and services. The figure 1 depicts the exact similarities between them. Virtualization is the main approach towards grids.

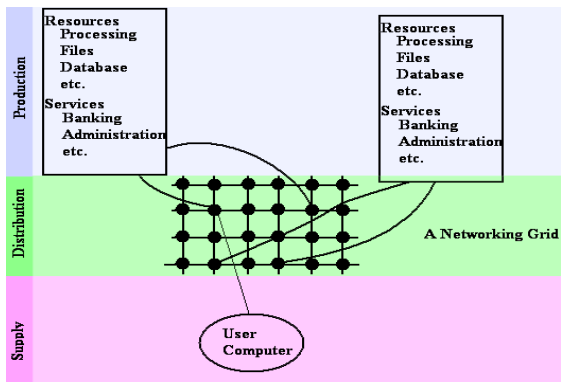

Figure 1. A Compute Grid

The Virtual organization (VO) based grids are those where services fall out of organizational boundaries, and multiple organizations combined render services to users. The user connects to a discovery node nearby it and gets to a service node in order to get service. Figure 2 shows a cloud containing a set of valid nodes ready to provide service. The service oriented grid middleware for e-learning as suggested in [1] needs to be furthered in aspects of e-governance.

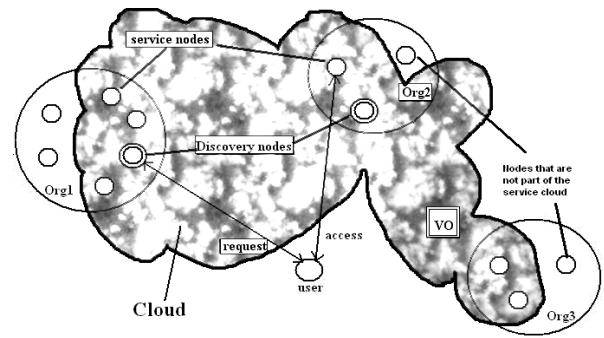

Figure 2. A Virtual Organization involving many organizations together (cloud).

Though many governments have already initiated in the aspects of computerization of certain services but the need of the time is to have an integrated approach to all types of government services under the framework of more customizable, distributed and scalable system, that is, obviously grids or clouds.

II. GOVERNMENT SERVICES AND SERVICE ORIENTED GRID / CLOUD APPROACH

As depicted in the figure 3 almost all government services functionally consists of 4 layers as far as manual systems are concerned. Without applying much change in the traditional levels of operation, a grid system can be set up which adds up more features like quick processing, transparency, reliable and above all accessible to common man.

The political or regulatory body consists of the state and the highest level officials (mostly politicians). The enforcement body will cover all districts and will consist of officials appointed by the state. The servicing body will cover divisions as-well as sub divisions. This body will be the actual face of the government operations and will interact with the masses directly.

The users are the citizen of the state viewed as part of a particular division or sub division. The grid system if incorporated will deem both government officials and citizen as its users as all of them will be using one or other services / applications available on the grid system (not the manual system). Only the authorization levels or state [8] will be different. An example in this regard is well depicted in [8] where a prototype grid on CMMS authorization model is presented for a state police services.

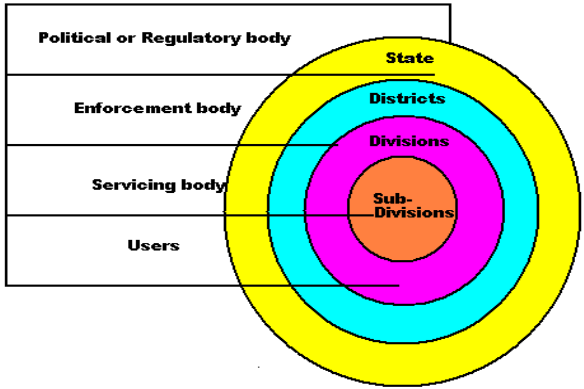

Figure 3. Generic functional structure of any government service.

The Grid or Cloud will consist of major 3 types of nodes namely:

1. Monitoring node.
2. Controlling node.
3. Service node.

The figures 4, 5 depict a functional structure of the grid/cloud. The functionality of monitoring node might be that of assigning roles, changing states [8], changing/creating/approving certificates etc. The monitoring node will have the major task of acting as discovery nodes, establish synchronization among service nodes, managing scaling of service nodes, imposing states [8] etc. The service nodes will be the actual nodes at various sites which will be operated by the service delivery officials who will use the end software to update records and perform transactions. The users will be accessing the grid through peer grid softwares or may be via internet.

III. INFRASTRUCTURAL AND IMPLEMENTATION ISSUES

We can broadly categorize the cloud e-governance implementation issues into following three factors:

A. Infrastructure or backbone development.
B. Software implementation.
C. Ensuring usability.

*A. Infrastructure or backbone development*

The grid/cloud based e-governance at its peak would require a well made and managed infrastructure throughout the province and beyond. The possibilities of having an eco-grid [2] for optimum resource utilization and at a economic mode can be framed similarly as in [2].

Structurally the clouds will consist of nodes servicing at remote frontiers. The users might well be accessing from long distances. The service nodes may need to synchronize in many cases and that too quite frequently. The controlling nodes of these service nodes might well be at some well defined establishments and may be rendering services along with control. The database synchronization and maintenance will be both at node level as well as grid/cloud level.

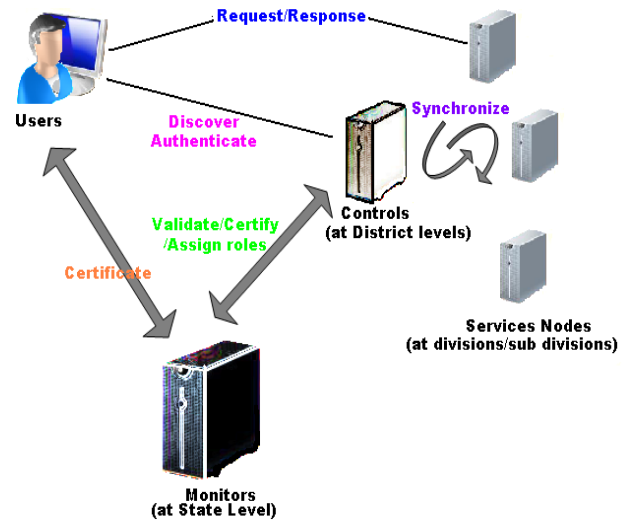

Figure 4. Role players in the Grid/Cloud.

*B. Software implementation*

Among many issues pertaining to any grid establishment, the factors of concurrency, security, and performance are major ones. The OGSI, WSRF, GSI [10] standards do take care of almost completely. The factor of authorization is vital in grid services and has to be taken care of separately than just depending on standards (GSI) alone. The CMMS model [8] suggests an authorization procedure especially for peer grid services. The model is mainly designed keeping in mind the governmental services.

*Service oriented architecture (SOA)*: The e-governance should be implemented over SOA As rightfully acknowledged in [3], the SOA brings exceptionally component based development with Enterprise Application Integration (EAI). Also, the fact of having peer to peer SOA (SOPs [4]) is more suitable with the general model presented for authorization i.e. CMMS [8].

The establishing of the entire infrastructure for internet based clouds might require sincere and serious effort from the state but once achieved can make a comfortable environment of governance and helps the governments to have better understanding of the need and demand of the polity as well as the mass.

The State Wide Area Networks (SWAN) [12] plan with the Indian government is a vital step towards achieving fruitful e-governance. The infrastructure will help in providing a backbone for an e-governance cloud for efficient governance of the mass.

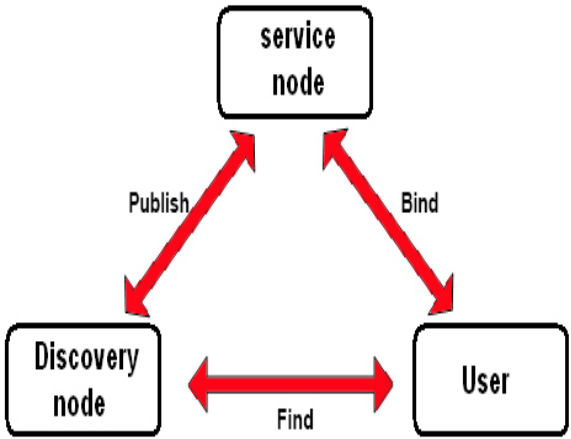

Figure 5. Basic service oriented architecture over grids.

Every grid applications has to rely upon a certification procedure that is robust, well defined and managed by a reliable third party. In case of using the CMMS model the third party will be the highest body in the state. However, since giving the keys in the hands of regulatory bodies may invite digital corruption, that is, corruption related to manipulation of digital signatures/certificates. Giving a secret and uniform identification numbers to all citizens would be required in order to have a corruption free and hassle free grid based e-governance application. The Nilekani's UID project [9, 11] in India is a step forward to realize the dream of services oriented grids for e-governance.

*C. Ensuring usability*

Ensuring usability is of prime concern as far as respective governments are concerned. A joint impact assessment study [5] was initiated in January 2006 by Indian Institute of Management and e-governance practice group of World Bank, Washington DC. The study was to define a framework and methodology for impact assessment of e-government projects from 3 states in India and 2 projects from Chile. The primary objective of this study was to measure the impact of computerization in selected service delivery projects. Though all the projects under assessment are currently for servicing urban clients, a promising fact comes out of this is, if the population is computer literate and has an infrastructural setup then gradually but steadily the manual delivery service is bound to get abolished.

It is quite certain that there is a huge mandate for the computerized systems. As for example KAVERI [4] was opened to service stamp duty and registration sector was started in December 2003. In 2000-01 when the system was manual only 0.63 million properties were registered. In 2005-06, with KAVERI system 1.02 million properties were registered representing an annual growth of 10.27%. Nearly 98% of respondents preferred the computerized system over the manual system because the time and cost of getting the service for clients has come down significantly. The survey also pointed out the major factors that the citizen demand for certain services. They prepared a chart by interviewing a set of citizen for their most desired factors for certain services like land registration, procurement, taxation etc. The table I is the original table revealing top four most desired attributes in particular software services. By having a closer look into the table we can easily figure out that major factors that are desired are improved governance, Quality and transaction efficiency in any services rendered through online or computerized system. Same are the prime factors must be taken care of on a servicing grid for e-government services. Among the three most desired attribute, the quality plays a major portion. Thus, the designers of the software over the grid infrastructure have to be careful that the quality factors are not compromised with.

| Project | Attribute 1 | Attribute 2 | Attribute 3 | Attribute 4 |
|---|---|---|---|---|
| **KAVERI** | <u>Less Corruption</u> | <u>Greater Transparency</u> | *Error free transaction* | **Less waiting time** |
| **Khajane – DDO** | *Simplicity of procedure* | *Convenient time scheduled* | *Friendly attitude of officers* | *Error free transaction* |
| **Khajane – Payee** | **No delay in transaction** | *Convenient time scheduled* | *Good location* | *error free transaction* |
| **eProcurement** | <u>No corruption</u> | *Easy access* | *Equal opportunity to all* | *No need to visit Government Office* |
| **eSeva** | **Less time and effort required** | **Less waiting time** | *Convenient time scheduled* | *Equal opportunity to all* |
| **Checkpost** | **No delay in transaction** | *Error free receipt* | *Error free transaction* | *Proper queue system* |
| **Legend:** Underline - Improve Governance; Bold - Transaction Efficiency; Italics - Quality | | | | |

TABLE I. TOP FOUR ATTRIBUTES DESIRED IN CERTAIN COMPUTERIZED SERVICES SURVEYED IN INDIA.

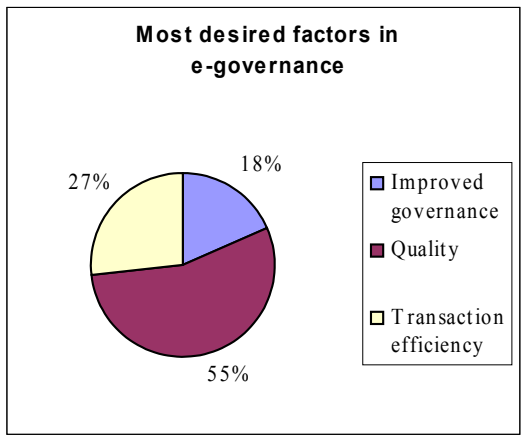

Figure 6. Percentage desire for generalized attributes in an e-governance software.

Major quality factors are: error free receipt, easy access, equal opportunity to all, proper queue system etc. Grid itself is a big solution for easy access. The certification authority and the authentication procedures have to be tight enough so that receipts are error free and can be regenerated if demanded.

In most of the cases the citizen users have always appreciated the online services and are sure to appreciate the fast, robust and reliable services on the grid infrastructures. Coming to verse with any new system is not difficult for any one provided a proper introduction and helping hand is extended. Nilekani [6] cites one instance after another of rural village dwellers, farmers, taxpayers, and others who quickly grasp what computers and Internet access can do for them. Whether it's the chance to learn English, check crop prices, or pay a utility bill, Indians at all levels come to depend on the computer once it's introduced. (The hard thing is persuading agencies and local officials to install systems that undercut their power as gatekeepers.) And we've all heard of the Hole in the Wall Project, where Indian kids in slums come to enjoy and figure out how to use computers with little or no adult help. This is true for any country and not only India.

## IV. Conclusion

Service oriented grids/clouds are fast emerging these days. In this paper we discussed how a service oriented grid could be utilized to provide useful e-governance. Government services can become more reliable and transparent. Services oriented grids can be realized using the SOA and CMMS models. Few primary layouts were discussed by showing the similarities between government structures and the suggested CMMS model. The issues pertaining to the realization of grid based e-governance were discussed emphasizing the points like implementation, usability and infrastructure. Current governments should take due steps to build a favorable infrastructure for grids. The citizens have to be empowered by steps like UID etc. sustained efforts from both government and citizens could help build a realistic service oriented grid that could benefit even the most common citizen of the state. The paper is intended to bring more light into this direction so that more and more intellect could be driven towards it. More sustained efforts are required to develop a complete and more concrete design of the service oriented grids for useful e-governance.

## V. References

[1] Wang GuiLing, Li YuShun, Yang ShengWen, Miao ChunYu, Xu Jun, Shi MeiLin, "Service-Oriented Grid Architecture and Middleware Technologies for Collaborative E-Learning," scc, vol. 2, pp.67-74, 2005 IEEE International Conference on Services Computing (SCC'05) Vol-2, 2005

[2] Rajkumar Buyya, David Abramson, Jonathan Giddy, "A Case for Economy Grid Architecture for Service Oriented Grid Computing", Proceedings of the 10th Heterogeneous Computing Workshop HCW 2001 (Workshop 1) - Volume 2 20083.1, 2001

[3] Alexander Sterff, "Analysis Of Service-Oriented Architectures From A Business And An It Perspective", Master's Thesis, 2006

[4] KAVERI - Karnataka Valuation And e-Registration , www.karigr.org

[5] Impact Assessment Study Of E-Government Projects In India, Center For e-Governance, Indian Institute Of Management, Ahmedabad, Department Of Information Technology, Government Of India, New Delhi, January 2007, http://www.iimahd.ernet.in/egov/documents/impact-assessment-study-dit.pdf

[6] Andy Oram, "Computerization in Nilekani's Imagining India", http://radar.oreilly.com/2009/09/computerization-in-nilekanis-i.html, as viewed on date 11-06-10

[7] "Structure of Indian Government" http://www.atatwork.org/content/download/427/2763/file/Attachment%2011%20Structure%20of%20Indian%20Government.pdf

[8] Ajay Prasad, A.K. Sharma, S.S. Verma, "Certification Authority Monitored Multilevel and Stateful Policy Based Authorization in Services Oriented Grids", International Journal of Security, (IJS), Volume (3) : Issue(4):48-64, 2009

[9] Creation of position of the Chairperson UID Authority of India, http://pib.nic.in/release/release.asp?relid=49370, as on 12.30 pm, 05/07/2010.

[10] Bart Jacob, Michael Brown, Kentaro Fukui, Nihar Trivedi. Introduction to Grid Computing, ibm.com/redbooksibm.com/redbooks. IBM Corporation 2005 Dec 2005.

[11] Nilekani takes charge, TNN, Jul 24, 2009, 05.06am IST, http://timesofindia.indiatimes.com/articleshow/4812763.cms, as on 12.30 pm, 05/07/2010

[12] State Wide Area Network (SWAN), http://www.nisg.org/kchome.php?page=9, as on 12.30 pm, 05/07/2010

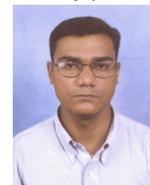

**Mr. Ajay Prasad, Assistant Professor, Dept of CSE, Sir Padampat Singhania University** did his M.Tech. in CSE from MITS University, Laxmangarh, following an MCA Degree from Nagpur University. He did his bachelors (B.Sc.) from Osmania University, Hyderabad. He has been rendering his services in Academic as well as s/w development fields since last 12 years and has been mentoring lots of project works ranging from academics to real time. He has contributed in the fields of security in Grid/Cloud and adhoc networks and security in social networks and he has been researching in probabilistic approaches in data mining techniques.  He is Life member of ISTE. Currently he is pursuing his PhD from SPSU.

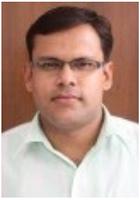

**Mr. Sandeep Chaurasia, Lecturer, Dept of CSE, Sir Padampat Singhania University** did his M.Tech. in CSE from Devi Ahilya University, Indore. He did his bachelors (B.E) from Rajasthan University, Jaipur. He has in academics and contributing as well in s/w development fields since last 3 years. He has contributed in the fields of application development in .NET Technology and software testing and he has been researching in Management Information Systems and knowledge management in business organization with quantification of knowledge management. Currently he is pursuing his PhD from SPSU.

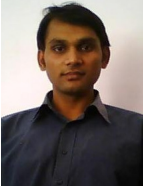

**Mr. Arjun Singh, Senior Lecturer in Sir Padampat Singhania University.** He did his M.Tech Degree with specialization in networking from VIT Vellore. His specialization is mainly in the Network Security and cryptography, wireless Grid Computing, Computer Network, Network design. Prior to his teaching assignment, he served in IBM & Capgemini pvt. Ltd. as a software engineer.

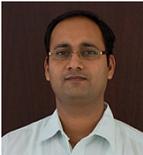

**Mr. Deepak Gour, Sr. Lecturer – Dept. of CSE, Sir Padampat Singhania University**, Udaipur completed his B.Sc. in Computer Science in 1998 & Master in Computer Application in 2001. Currently he is Perusing Ph.D. from Department of Computer Science, Mohan Lal Sukhadia University, Udaipur. His research area is in ASIP Design Space Exploration. His Area of Specialization is in Embedded Systems and his Research interest lies in Application Specific Instruction set Processor.